\documentclass[a4paper,fleqn, review]{cas-sc}

\usepackage[utf8]{inputenc}
\usepackage{graphicx}
\usepackage{flafter}

\usepackage{amsmath,amssymb,amstext,amsthm,amsfonts}
\usepackage{braket}
\usepackage{bm}
\usepackage{float}
\usepackage{color}
\usepackage{subfigure}
\usepackage{blindtext}
\usepackage{marginnote}
\usepackage{ulem}

\usepackage[numbers, sort&compress]{natbib}

\def\tsc#1{\csdef{#1}{\textsc{\lowercase{#1}}\xspace}}
\tsc{WGM}
\tsc{QE}

\begin{document}
\let\WriteBookmarks\relax
\def\floatpagepagefraction{1}
\def\textpagefraction{.001}

\shorttitle{Singular flat bands in the modified Haldane-Dice model}    

\shortauthors{Alexander Filusch and Holger Fehske}  

\title [mode = title]{Singular flat bands in the modified Haldane-Dice model}


\author[1]{Alexander Filusch}[orcid=0000-0003-1581-9388]

\cormark[1]

\credit{Conceptualization, Formal analysis, Investigation, Writing - Original Draft}

\ead{alexander.filusch@uni-greifswald.de}

\author[1,2]{Holger Fehske}[orcid=0000-0003-2146-8203]

% Footnote of the second author

% Email id of the second author
\ead{fehske@physik.uni-greifswald.de}


\credit{Conceptualization, Writing - Reviewing and Editing, Supervision}

\affiliation[1]{organization={Institute of Physics, University Greifswald},
	addressline={}, 
	city={Greifswald},
	%          citysep={}, % Uncomment if no comma needed between city and postcode
	postcode={17489}, 
	country={Germany}}

\affiliation[2]{organization={Erlangen National High Performance Computing Center},
	addressline={}, 
	city={Erlangen},
	%          citysep={}, % Uncomment if no comma needed between city and postcode
	postcode={91058}, 
	country={Germany}}
\cortext[1	]{Corresponding author}



\begin{abstract}
	Flat bands can be divided into singular and non-singular ones according to the behavior of their Bloch wave function around band-crossing points in momentum space.
	We analyze the flat band in the Dice model, which can be tuned by a uniaxial strain in the zigzag direction and a Haldane-type next-nearest neighbor interaction, and derive the topological phase diagram of the modified Haldane-Dice model to obtain all band-gap closings with the central band. When the central band is flat, we determine its compact localized state and classify its behavior at all band-touching points by means of the Hilbert-Schmidt quantum distance.
	We find that the flat band remains singular for all band-touching points (topological phase transitions) with a maximal quantum distance and give expressions for the resulting non-contractible loop states on the real-space torus.
\end{abstract}


% Keywords

\maketitle

% Main text

\section{Introduction}
\label{Intro}

Flat bands are dispersionless, i.e., have constant energy throughout the entire Brillouin zone, implying that their charge carriers have zero group velocity (or infinite effective mass) and a diverging density of states~\cite{MHMRD12, LAF18, RY21}. 
Because of the quenched kinetic energy, the charge carriers are then dominated by the electron-electron interaction, making flat-band systems an ideal platform to study the fractional quantum Hall effect~\cite{WGSY12, PRS13}, superconductivity~\cite{HV11, BDEY20}, and Wigner crystallization~\cite{WBBS07}.
Experimentally, various systems with perfectly flat energy bands have been accomplished in, e.g.,  optical~\cite{HAM13, TOINNT15} and photonic lattices~\cite{VCMRMWSM15, MSCGOAT15, XHSZTC16} or metamaterials~\cite{NONK12}.

%CLS 
The perfectly localized eigenstates of the flat band can be used for the classification of flat bands and their generators~\cite{R17, MAPGF17, MFA19, MAF21}. Due to the destructive interference, these compact localized states (CLS) have nonzero amplitude only inside a finite real-space region and remain intact under time-evolution~\cite{Su86, AAM96, MAPGF17, RY19}. The basis of an isolated flat band in a lattice of $N$ unit cells is spanned by the $N$ linearly independent CLS generated by lattice translations. 

In the class of singular flat bands, however, this set is linearly dependent on a real-space torus (periodic boundaries) because dispersive bands cross the flat band in the Brillouin zone. Therefore the set of all CLS must be complemented by so-called non-contractible loop states (NLS), which extend around the torus along one spatial direction while being localized along the other and cannot be smoothly deformed by the CLS~\cite{BWB08, RY19}. The existence of CLS and NLS have been demonstrated for photonic Lieb and Kagome lattices~\cite{VCMRMWSM15, MSCGOAT15, XRXSLLTHSXLFC18, MRTXWZXSHLYLC20} with diffraction-free image transmission~\cite{XHSZTC16}.
Interestingly, the band crossing point is directly related to an immovable singularity in the Bloch wave function of the flat band in momentum space~\cite{RY19, RY21} and is measured by the maximum Hilbert-Schmidt quantum distance~\cite{OCPR22}. This has direct consequences for systems with quadratic band-touching points, giving rise to anomalous Landau levels~\cite{RKY20} and boundary modes through the bulk-interface correspondence~\cite{OCPR22}. The full quantum metric tensor, whose real part is the quantum distance, has recently attracted a lot of attention in the context of superconductivity~\cite{KRCB20, APSHB22, HACBT22},  the geometric orbital susceptibility~\cite{PRFM16}, and the dc linear conductivity~\cite{MH22}.

\begin{figure*}[]
	\centering
	\includegraphics[scale=0.9]{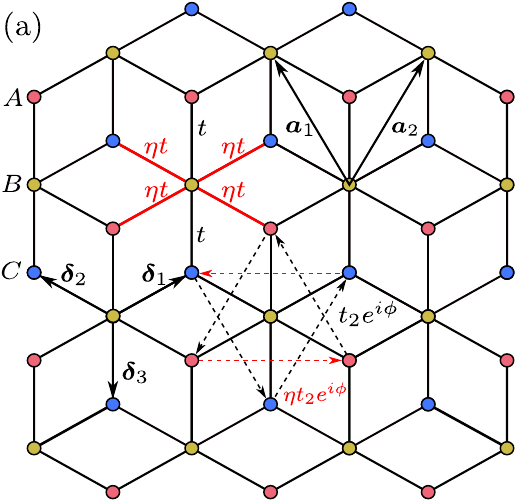}
	\includegraphics[scale=0.9]{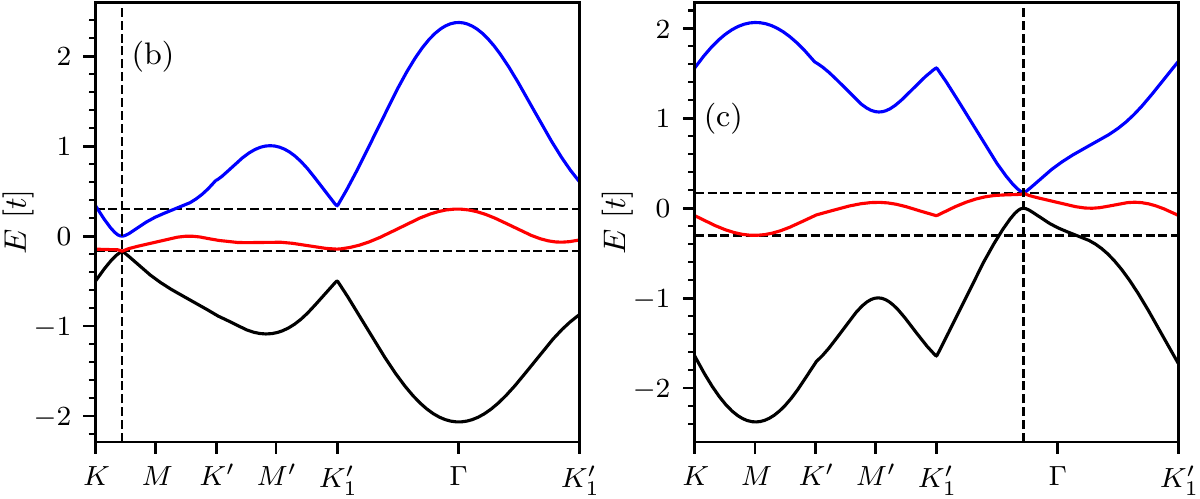}
	\caption{(a) Modified Haldane-Dice model with basis $\{A,\, B,\, C\}$ and Bravais lattice vectors $\bm{a}_1=(-\sqrt{3}/2, 3/2)$ and $\bm{a}_2=(\sqrt{3}/2, 3/2)$. The NN vectors connecting the $B$ sites with $C$ sites are given by $\bm{\delta}_1 = (\sqrt{3},1)/2$, $\bm{\delta}_2=(-\sqrt{3},1)/2$ and $\bm{\delta}_3=(0,-1)$. (b),(c) Band structure along the high-symmetry directions with a Haldane flux $\phi=0.3\pi$  at the topological phase transition $m_c(\phi, \eta)$ of Equation~\eqref{eq:phase_boundary} between  two valence-gapped phases for $\eta=0.6$ (a) or between two conduction-gapped phases $\eta=-0.6$ (b). The vertical dashed line denotes the band-crossing point $\mathbf{k}_c$ in the Brillouin zone. Horizontal dashed lines denote the maximum and minimum of the central band. }
	\label{fig1}
\end{figure*}

Historically, the Dice (or $\mathcal{T}_3$) lattice ~\cite{Su86,VMD98,VBDM01}, a honeycomb lattice with an extra atom $C$ placed in the center of each hexagon and coupled to only one of the sublattices $A$ or $B$ [cf. Figure~\ref{fig1}(a)], is perhaps the earliest example of a flat-band system.
The spectrum of the Dice model consists of a strictly flat band, crossing the respective Dirac cones around the $K$ and $K'$ points inherited from graphene's band structure. 

The chiral flat band has been shown to be stable against perturbations of the transfer amplitude like strain, magnetic fields, or boundary conditions due to the bipartite nature of the lattice and plays only a secondary role for the single-particle transport properties~\cite{Su86, BG16, RAF17, OGG18, BO19, MCAF22, FBSWH21, FF22}, but its geometry and composition have not yet been discussed in detail.

In this paper, we address these issues, analyzing the Dice model under a uniaxial strain along the zigzag orientation with a Haldane-type next-nearest neighbor (NNN) interaction. The uniaxial strain breaks the threefold rotational ($\mathcal{C}_3$) symmetry and trimerizes the model along the $y$-axis. 
The Haldane term breaks time-reversal symmetry without a net flux and drives a topological phase transition to Chern insulator in the Dice model characterized by the Chern number $c=\pm 2$~\cite{Ha91, DKPG20, MB23}. 
By studying the quantum distance in the vicinity of the band-crossing points at the phase transition of the suchlike modified Haldane-Dice model, we show that the singularity of the flat band is related to its geometry. We confirm this by showing that the set of the CLS is linearly dependent and presenting expressions for the NLS.

\section{Modified Haldane-Dice model and phase diagram}

We start by introducing the Hamiltonian of the modified Haldane-Dice model (MHDM),
\begin{align}
H_\eta &=\left[ \frac{1}{\sqrt{2}}\sum \limits_{\langle i,j\rangle} t_{ij} c_i^\dagger c_j + \sum \limits_{\langle\langle i,j\rangle\rangle}t_{2, ij}e^{-i \nu_{ij}\phi} c_i^\dagger c_j + \text{h.c.} \right] \nonumber \\& \quad + m \sum \limits \epsilon_i c_i^\dagger c_j\,, \label{eq:H_TB}
\end{align}
obtained by applying uniaxial strain along the $x$ direction.
Here, the nearest-neighbor (NN) transfer amplitudes are  $t_{ij}=t$ along the $\pm\bm{\delta}_3$ direction and $t_{ij}=\eta t$ along the $\pm \bm{\delta}_1$ and $\pm\bm{\delta}_2$ direction [cf. Figure~\ref{fig1}(a)].  The NNN couplings acquire an alternating magnetic phase accounted for by the sign $\nu_{ij}$=1 $(-1)$ when the hopping between $A$-$A$ and $C$-$C$ is clockwise (anti-clockwise). The NNN hoppings in the $B$ sublattice are suppressed~\cite{DKPG20}. The value of the NNN transfer is $t_{2, ij}=\eta t_2$ $[t_2]$ along $\bm{a}_3=\bm{a}_1+\bm{a}_2$ [$\bm{a}_1$ and $\bm{a}_2$]. By making the NN and NNN couplings unequal, the $\mathcal{C}_3$ rotational symmetry is broken. The inversion-symmetry breaking on-site potential (Semenoff mass) is denoted by $m$, where $\epsilon_i=$\,1, 0, $-1$ when $i$ refers to the $A$, $B$ or $C$ sublattice, respectively. Throughout this work, we consider $\eta\in [-1,1]$.
\begin{figure*}[]
	\centering
	\includegraphics[scale=0.9]{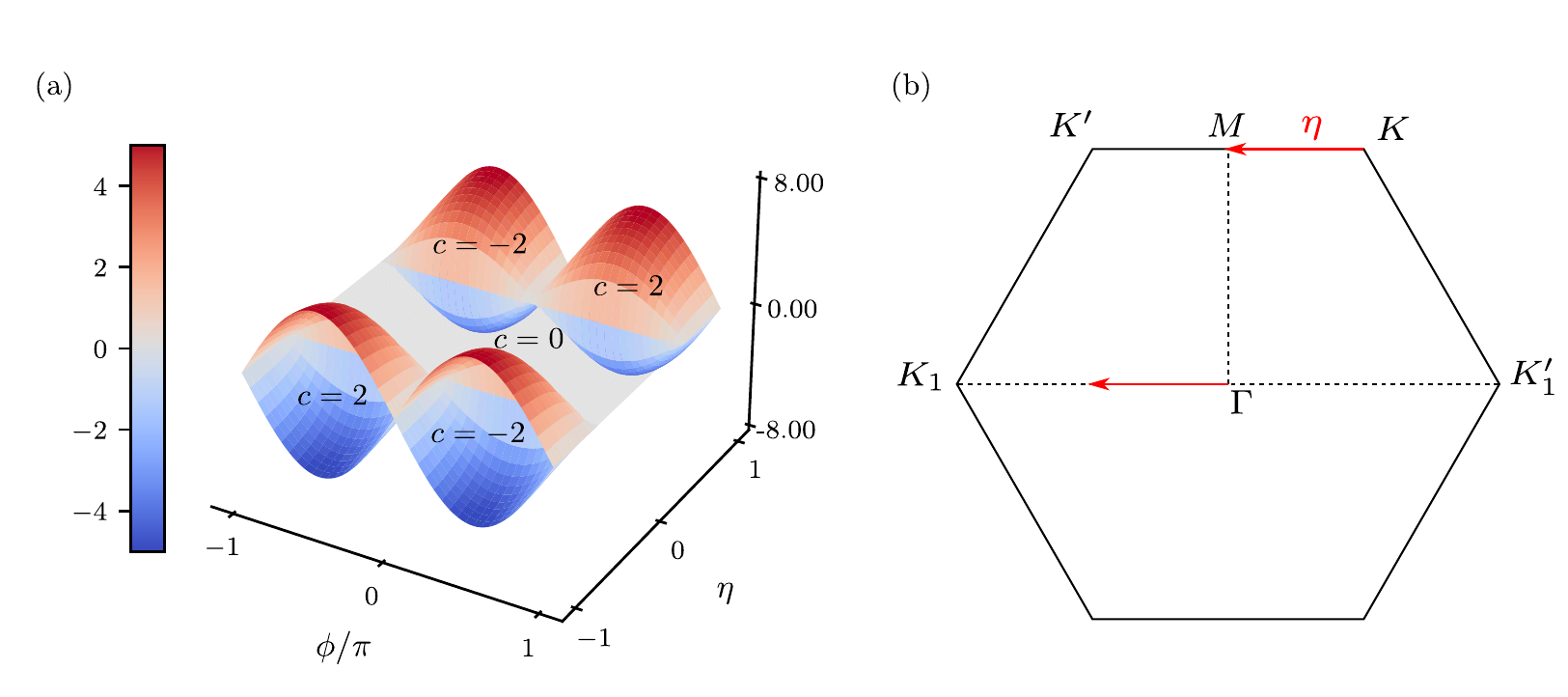}

	\caption{(a) Phase diagram of the modified Haldane-Dice model. The Chern numbers are displayed for the valence or valence and flat band. (b) Variation of the band crossing in the Brillouin zone for positive $m_c$ in dependence on $\eta$.}
	\label{fig2}
\end{figure*}
In momentum space, the Hamiltonian becomes
\begin{align}
H_\eta(\mathbf{k})&=2 t_2 h_0^\eta(\mathbf{k})\cos(\phi) S_0 + \left[m-2t_2 h_z^\eta(\mathbf{k})\sin (\phi) \right]S_z \nonumber \\& \quad+ t f_\eta(\mathbf{k}) S_x \, ,\label{eq:H_k}
\end{align}
where 
\begin{align}
f_\eta(\mathbf{k}) &= \eta \left( e^{i \mathbf{k} \cdot \mathbf{a}_1} + e^{i \mathbf{k} \cdot \mathbf{a}_2}\right) +1 \, , \nonumber \\
h^\eta_0(\mathbf{k}) &= \eta \cos(\mathbf{k}\cdot \bm{a}_3)+\cos(\mathbf{k}\cdot \bm{a}_2)+ \cos(\mathbf{k}\cdot \bm{a}_1)\, , \nonumber \\ 
h^\eta_z(\mathbf{k}) &= \eta \sin(\mathbf{k}\cdot \bm{a}_3)+\sin(\mathbf{k}\cdot \bm{a}_2)-\sin(\mathbf{k}\cdot \bm{a}_1)\, .
\end{align}
In Equation~\eqref{eq:H_k}, the matrices $S_x$, $S_z$ are the usual spin-1 matrices, and $S_0 =\text{diag}(1,\,0,\, 1)$.

To determine the topological properties of the MHDM, we consider the Chern numbers for each band $n$~\cite{TKNN82},
\begin{align}
c_n = \frac{1}{2\pi} \iint_{\text{BZ}} \Omega^{(n)}(k_x, k_y) \, \mathrm{d}k_x\,\mathrm{d}k_y,
\end{align}
where $\Omega(k_x, k_y)= \left[\nabla_{\mathbf{k}}\times \mathbf{A}^{(n)}\right]_z$ is the Berry curvature, $\mathbf{A}^{(n)}=\langle u_n(\mathbf{k})|i\nabla_\mathbf{k} | u_n(\mathbf{k})\rangle$ is the Berry connection, and $|u_n(\mathbf{k})\rangle$ denotes a Bloch state.

Although the gap-closings are identical to that of the modified Haldane model of graphene~\cite{WZLB21}, the situation is slightly different due to the three-band character. Here, the finite NNN transfer generally distorts the central band, only a Haldane flux of $\phi_c=(2n+1)\pi/2$ with integer $n$ restores the flatness~\cite{DKPG20}. In this case, the flat band has always $c=0$~\cite{CMST14}.
Therefore, the band structure of the MHDM differentiates between the valence-band gapped, conduction-band gapped, and all-band gapped phases. For example, the valence-gapped phase has a direct gap between the central band and the valence band and an indirect overlap between the central band and the conduction band. Figure~\ref{fig1}(b) [(c)] displays the band structure of the MHDM at the phase boundary between two valence-band gapped phases [conduction- band gapped phases]. In Figure~\ref{fig2}, we show the Chern numbers obtained by the discretized Brillouin zone method of Fukui \textit{et al.}~\cite{FHS05} for the valence band in the valence-band and all-band gapped phases or for valence and flat band in the conduction-band gapped phase. For Fermi energies inside the indirect gap, the overlapping central and conduction/valence band gives the system a metallic character [cf. Figure~\ref{fig1}(b) and (c)].

In the unmodified case ($\eta=1$), the system undergoes a topological phase transition from a Chern insulator with $c=\pm 2$ to a trivial insulator when $m_c=\pm 3\sqrt{3} \sin \phi$~\cite{DKPG20}. The band-gap closing occurs at the $K$ or $K'$ points since the Hamiltonian in Equation~\eqref{eq:H_k} is $\mathcal{C}_3$ symmetric~\cite{DKPG20}.
For $\eta \in [-1,\,-0.5) \cup (0.5, 1]$, we find that the surface
\begin{align}
m_c(\phi, \eta)= \pm 3 t_2 \sin \phi \sqrt{4-\frac{1}{\eta^2}} \label{eq:phase_boundary}
\end{align}
describes the gap-closing/reopening in the parameter space, separating the Chern-insulating phase of the MHDM with $c=\pm 2 $ from a trivial insulator.

To understand the gap opening in terms of the strain, we monitor the position of the band crossings in the Brillouin zone. On the surface given by Equation~\eqref{eq:phase_boundary}, the MHDM has one or two band-crossing points depending on the time-reversal symmetry. For $\phi=\phi_c$, the system is a semimetal that hosts one pseudospin-1 Dirac node in the Brillouin zone. The presence of time-reversal symmetry ($\phi=0$, $\pi$) enforces the appearance of two band-crossing points. In all cases, the position of the Dirac nodes depends merely on $\eta$. We find that the conduction or valence band touches the central band at $\mathbf{k}_c = \left[\text{sgn}(m_c) k_x^\eta, k_y^\eta\right]$, where
\begin{align}
k_x^\eta = \frac{2}{\sqrt{3}} \arccos\frac{1}{2\eta}- \Theta\left(\frac{1}{2}-\eta\right)\frac{2\pi}{\sqrt{3}} 
\end{align}
and $k_y^\eta =2\pi/3$ for $\eta\geq 1/2$ and $k_y^\eta =0$ for $\eta\leq -1/2$, respectively. The position of the band-crossings $\mathbf{k}_c$ is located on the symmetry-invariant lines $K' - M - K$ and $K_1 - \Gamma - K_1'$ [cf. Figure~\ref{fig1}(b) and (c)], and depends on the $\eta$ parameter as illustrated in Figure~\ref{fig2}(b) for positive $m_c$.
By reducing $\eta\rightarrow 1/2$, the $\mathcal{C}_3$ symmetry is broken and the Dirac node is no longer pinned to the symmetry point $K$ and moves to $M$. At $\eta=1/2$, the two Dirac nodes merge at the $M$ point and gap out. The MHDM remains gapped for smaller $\eta$ due to the trimerization. For $\eta=-1/2$, the Dirac node reappears at the $\Gamma$ point and moves towards the $K_1$ point when $\eta\rightarrow -1$.  
% What do we find regarding the eta-dependency of the CG AG VG phases? 

\section{Singular band touchings}

To determine whether the flat band is singular, we evaluate the Hilbert-Schmidt quantum distance~\cite{DMMW00} 
\begin{align}
d(\mathbf{k}_1, \mathbf{k}_2) = 1 - |\langle\Psi_\mathrm{FB}(\mathbf{k}_1)|\Psi_\mathrm{FB}(\mathbf{k}_2)\rangle|^2 \label{eq:q_dist}
\end{align}
of the flat-band eigenstates in the vicinity of the band-touching points. Typically, the quantum distance vanishes for two eigenstates that are arbitrarily close to each other. In the singular case, however, this quantity is nonzero for $|\mathbf{k}_1 - \mathbf{k}_2|\rightarrow 0$~\cite{RKY20}. The maximum value of the quantum distance $d_\mathrm{max}$ has been recently taken as a measure of the singularity of the band-crossing point, where any $d_\mathrm{max}>0$ implies the linear dependence of the CLS and the existence of NLS~\cite{RKY20, RY21, OCPR22}.

\begin{figure}
	\centering
	\includegraphics[scale=1]{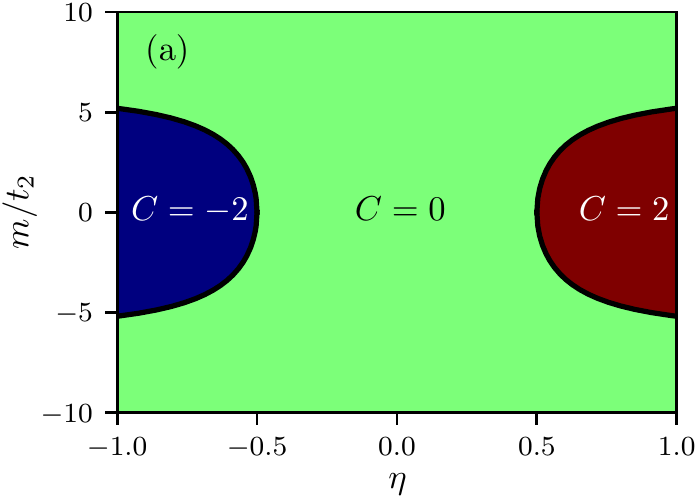}
	\includegraphics[scale=1]{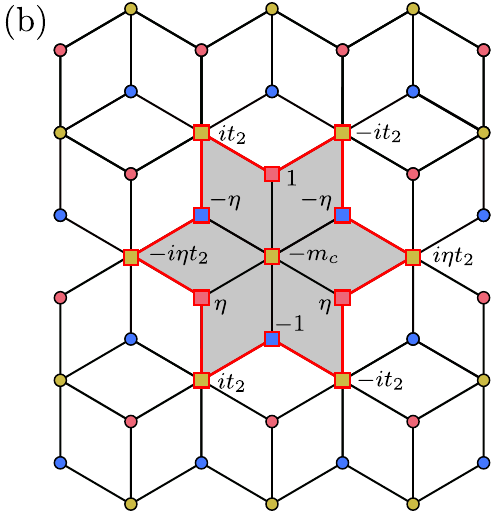}
	\caption{(a) Chern numbers in the Dirac semimetal phase ($\phi=\pi/2$) for the valence band. (b) Real-space distribution of the (unnormalized) compact localized state on the Dice lattice, where squares mark sites with non-vanishing amplitude bordering the gray region. The NNN couplings are not drawn for better visibility.}
	\label{fig3}
\end{figure}

Let us now inspect the geometric properties of MHDM when the central band is flat, i.e., $E=0$  [$\phi_n=(2n+1)\pi/2$] at the phase transitions curves $m=m_c(\eta, \phi_n)$  [cf. Figure~\ref{fig3}(a)]. In the following, we will show that a nonzero $d_\mathrm{max}$ also implies a singular flat band for Dirac semimetal phases. The eigenstate of the flat band in these cases is given by
\begin{align}
|\Psi_\mathrm{FB}(\mathbf{k})\rangle = \mathcal{N}_\mathbf{k}\begin{pmatrix}
f_\eta(\mathbf{k})  \\
-(m_c - 2t_2 h_z^\eta(\mathbf{k}))\\
- f_\eta(\mathbf{k})^*\\
\end{pmatrix},
\end{align}
where $\mathcal{N}_\mathbf{k}= \left\{2 |f_\eta(\mathbf{k})|^2+(m_c - h_z^\eta(\mathbf{k}))^2\right\}^{-1/2}$ is the normalization constant. 
% v2 qd
In general, the value of $d(\mathbf{k}_1, \mathbf{k}_2)$ depends on how we approach the singular point $\mathbf{k}_c$. Therefore we use the parameterization $\mathbf{k}_{1/2} =\mathbf{k}_c + \mathbf{q}_{1/2} (\cos \theta_{\mathbf{q}_{1/2}}, \sin \theta_{\mathbf{q}_{1/2}})$ and first consider $\eta \neq \pm 1/2$. In this case, we have $m_c -2t_2 h_z^\eta(\mathbf{k}_{1/2}) = 0$ in the vicinity of $\mathbf{k}_c$ and the quantum distance for small $q_{1/2}$ is given by
\begin{align}
d(\mathbf{k}_1, \mathbf{k}_2) = \sin^2 \left(\Theta_{\mathbf{k}_1} - \Theta_{\mathbf{k}_2} \right), \label{eq:q_dist_t2=0}
\end{align}
where $\Theta_\mathbf{k}= i\log f_\eta(\mathbf{k})/|f_\eta(\mathbf{k})|$. This is similar to neglecting the NNN interaction. However, at $\eta = \pm 1/2$, $m_c(\eta=\pm 1/2)=0$, the full wave function has to be used which affects the angular dependence but not its maximum value. 
In the limit $q_{1/2} \rightarrow 0$, we find the unit maximum quantum distance $d_\mathrm{max}=1$ for, e.g., $\theta_{\mathbf{q}_1}=\pi/2$ and $\theta_{\mathbf{q}_2}= (2j+1)\pi$ for integer $j$ and independent of $\eta$.

At this point, we like to note that in the closely related $\alpha$-$\mathcal{T}_3$ model~\cite{RMFPM14}, the parameter $\alpha$ interpolates between the honeycomb lattice ($\alpha=0$) with a decoupled flat band and the Dice lattice ($\alpha=1$)  rescaling the hoppings between $B$ and $C$ sites by $\alpha$. In this case, we find $d(\mathbf{k}_1, \mathrm{k}_2)=\sin^2(\Theta_{\mathbf{k}_1}-\Theta_{\mathbf{k}_2})\sin^2 2\varphi$ with $\tan \phi = \alpha$. Thereby, the resulting $d_\mathrm{max}=\sin^2 2\varphi$ is a direct measure for the interband coupling between the flat band and the Dirac nodes.

To confirm the predictions by the maximal quantum distance, we consider a (Dice) lattice of $N$ unit cells with $\phi_0=\pi/2$ and $m=m_c(\eta, \phi_0)$, i.e., the MHDM where the conduction and valence band touch the perfectly flat band at $\mathbf{k}_c$.
We construct the compact localized state of the flat-band states by combining all Bloch wave functions into new eigenfunctions,
\begin{align}
|\chi(\mathbf{R})\rangle =  \mathcal{N} \sum \limits_{\mathbf{k}\in \text{BZ}} \alpha_\mathbf{k}e^{-i\mathbf{k}\cdot \mathbf{R}} |\Psi_\mathrm{FB}(\mathbf{k})\rangle,
\end{align}
where $\mathcal{N}$ is a normalization constant and $\alpha_\mathbf{k}$ is a smooth function of the momentum. The Bloch eigenstate can be brought in the form
\begin{align}
|\Psi_\mathrm{FB}(\mathbf{k})\rangle = \frac{1}{\sqrt{N}} \sum \limits_{\mathbf{R}'} e^{i\mathbf{k}\cdot \mathbf{R}'}  u_{\mathbf{k},j} |\mathbf{R}',j \rangle,
\end{align}
where $u_{\mathbf{k},j}$ is the $j$th component of the normalized eigenvector of the flat band.
By choosing $\alpha_{\mathbf{k}} = \mathcal{N}_\mathbf{k}^{-1}$, $\alpha_\mathbf{k} |\Psi_\mathrm{FB}(\mathbf{k})\rangle$ is simply a sum of Bloch phases and we  directly obtain the CLS real-space distribution centered around $\mathbf{R}'$, 
\begin{align}
\langle \mathbf{R}', j | \chi(\mathbf{R})\rangle =\frac{1}{\sqrt{2+6\eta^2+4t_2^2+m_c^2}}  \cdot  \begin{pmatrix}
\eta \left[\delta_{\mathbf{R}'-\mathbf{R}+\bm{a}_1} +\delta_{\mathbf{R}'-\mathbf{R}+\bm{a}_2} \right] +\delta_{\mathbf{R}'-\mathbf{R}} \\
-m_c -i t_2 \sum \limits_{i=1}^3 \beta_i \left(\delta_{\mathbf{R}'-\mathbf{R}-\bm{a}_i}-\delta_{\mathbf{R}'-\mathbf{R}+\bm{a}_i}\right)\\
-\eta \left[\delta_{\mathbf{R}'-\mathbf{R}-\bm{a}_1} +\delta_{\mathbf{R}'-\mathbf{R}-\bm{a}_2} \right] -\delta_{\mathbf{R}'-\mathbf{R}} \\
\end{pmatrix}, \label{eq:CLS}
\end{align}
where the amplitudes are $\beta_1 = \eta$ and $\beta_{i\neq 1} = 1$. Figure~\ref{fig2}(a) displays the (unnormalized) CLS. Due to the NNN coupling, the CLS of the pristine Dice lattice~\cite{Su86} is extended by complex amplitudes on the outer $B$ atoms. The additional nonzero amplitude on the central $B$ atom compensates the sublattice-symmetry breaking Semenoff mass.
The CLS $|\chi(\mathbf{R})\rangle$ can be considered as the extreme limit of localized Wannier functions~\cite{R17}.

\begin{figure*}
	\includegraphics[scale=0.95]{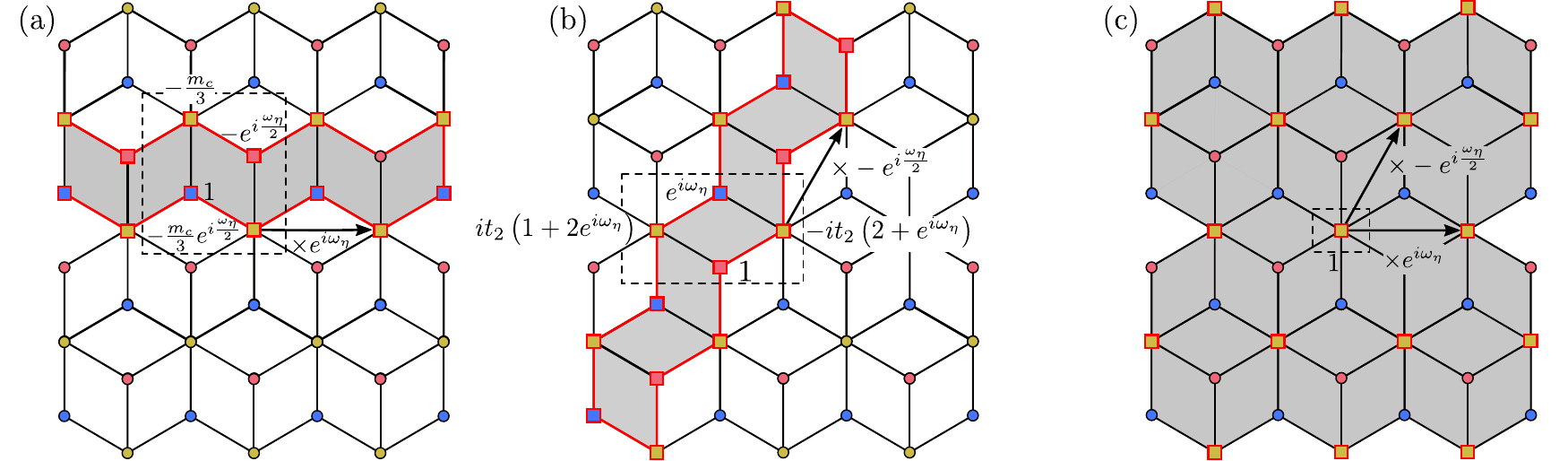}
	\caption{(a),(b) Non-contractible loop states winding along $\bm{a}_3$ and $\bm{a}_2$, respectively, around the torus. The phase factor is defined by $\omega_\eta =2\arccos(1/2\eta)$. All sites with nonzero amplitude are indicated by squares on the edges of the gray stripes. The values of the amplitudes are given for the unit plaquette (dashed box) and all other plaquettes can be constructed by shifting the unit plaquette via the phase factor along the arrow. (c) An extended state with nonzero amplitude on the $B$ sites in the gray region. The remaining amplitudes on the $B$ sites are obtained by shifting the nonzero amplitude in the dashed box by the phase factor along the two arrows.}
	\label{fig4}
\end{figure*}
For $N$ unit cells, one would expect $N$ eigenstates with zero energy from the flat band and two additional states stemming from the Dirac cone. In a system with periodic boundary conditions however, we can easily verify that  $\sum_n e^{i\mathbf{k}_c \mathbf{R}_n} |\chi(\mathbf{R}_n) = 0$, which renders the set spanned by all CLS linearly dependent. Thus, we are left with a set of ($N-1$) CLS, quite similar to the pristine case~\cite{BWB08}.
Therefore, we give expressions for the three missing zero-energy eigenstates in Figure~\ref{fig4} that complete the set. We find two NLS that extend along the $\bm{a}_3$ and $\mathbf{a}_2$ direction (or, equivalently, any two choices of the three $\bm{a}_i$), which are displayed in Figure~\ref{fig4}(a) and Figure~\ref{fig4}(b), respectively. Each non-contractible loop state consists of a unit plaquette with nonzero amplitude on its edge sites (dashed box). The full non-contractible loop state is then generated by shifting the plaquette and multiplying by a phase factor $\exp{\{i \mathbf{k}_c \cdot n\bm{a}_{3/2}\}}$. For the NLS along $\mathbf{a}_3$ and $\mathbf{a}_2$, this yields a phase factor of $e^{i\omega_\eta}$ and $-e^{i\omega_\eta/2}$, respectively. Additionally, we find an extended state in Figure~\ref{fig4}(c) with nonzero amplitudes only on the $B$ atoms. This state has been overlooked previously~\cite{BWB08}.

Finally, let us explicitly demonstrate that the NLS of Figure~\ref{fig4}(a) is an exact eigenstate with zero energy by using the picture of destructive interferences. First, we consider the $B$ site with weight $-m_c e^{i\omega_\eta/2}/3$ [cf. yellow square inside dashed box of Figure~\ref{fig4}(a)]. Since this $B$ site is only coupled to its NN [cf. Equation~\eqref{eq:H_TB}] and $H_\eta$ has no onsite potential on the $B$ sublattice, applying $H_\eta$  yields  $\eta+\eta e^{i\omega_\eta}-e^{i\omega_\eta/2} =0$ on this site. 
The remaining hoppings cancel each other, which can be seen by noting that $m_c = -3 i t_2 \eta (e^{i\omega_\eta}- e^{-i\omega_\eta}) = -3 i t_2 (e^{i\omega_\eta/2}-e^{-i\omega_\eta/2})$ and collecting the NN and NNN contributions. Due to translation symmetry, the NLS is then an eigenstate. Similarly, one can show that the NLS along $\bm{a}_2$ is also an eigenstate.

\section{Conclusions}
\label{Conclusions}
In summary, we studied the Haldane-Dice model with broken $\mathcal{C}_3$ symmetry due to different couplings between NN and NNN sites particularly with regard to its topological properties and the geometry of the flat band. In experiments, this could be realized by applying uniaxial strain to related systems. We found the same topological phase diagram as in the modified Haldane model based on graphene, albeit with an increased Chern number $c=\pm 2$.  From the phase diagram, we identified the parameters where the central band is dispersionless and determined the shape of the Bloch wave function and composition of the flat band.
In particular, we demonstrated that a nonzero maximum Hilbert-Schmidt quantum distance around the band-crossing point in the Dirac semimetal phase directly leads to a linearly dependent set of all compact-localized states in the torus geometry. 
In this case, we found two non-contractible loop states winding around the torus along the symmetry directions and one extended state on the torus that has not been reported so far. The tunability of the band-touching points and non-contractible loop states by strain might be advantageous for its detection in future experimental realizations of the modified Haldane-Dice model.
Accordingly, the full physical significance of the quantum metric tensor and quantum distance in this system was open.
In this regard, the $\alpha$-$\mathcal{T}_3$ model and its singular flat band is certainly an ideal platform to study geometric contributions to interband coupling effects arising from the $\alpha$-dependent quantum distance around the band-crossing points.

%------------------------------

\printcredits

\bibliographystyle{model1-num-names}
%\bibliographystyle{cas-model2-names}


\end{document}